\documentclass[aps,prd,english,showpacs,11pt]{revtex4-1}
\usepackage[latin1]{inputenc}
\usepackage{ucs}
\usepackage{amsmath}
\usepackage{amsfonts}
\usepackage{amssymb}
\usepackage{graphicx}
\usepackage{psfrag}
\usepackage{picinpar}
\usepackage{hyperref}

\newcommand{\rd}{{\rm d}}
\allowdisplaybreaks[1]
\newcommand{\be}{\begin{equation}}
\newcommand{\ee}{\end{equation}}
\newcommand{\ben}{\begin{eqnarray}}
\newcommand{\een}{\end{eqnarray}}
\newcommand{\bes}{\begin{subequations}}
\newcommand{\ees}{\end{subequations}}

\newcommand{\Frac}[2]{\frac{{\displaystyle #1}}{{\displaystyle #2}}}

\begin{document}
\title{Discussion on the energy content of the galactic dark matter
  Bose-Einstein condensate halo in the Thomas-Fermi approximation}

\author{J. C. C. de Souza} 
\email{jose.souza@ufabc.edu.br}
\author{M. O. C. Pires}
\email{marcelo.pires@ufabc.edu.br}

\affiliation{Centro de Ci\^{e}ncias Naturais e Humanas, Universidade
  Federal do ABC, Rua Santa Ad\'elia 166, 09210-170, Santo Andr\'{e}, SP,
  Brazil}

\begin{abstract}
We show that the galactic dark matter halo, considered composed of an
axionlike particles Bose-Einstein condensate \cite{pir12} trapped by a
self-graviting potential \cite{boh07}, may be stable in the
Thomas-Fermi approximation since appropriate choices for the dark
matter particle mass and scattering length are made. The demonstration
is performed by means of the calculation of the potential, kinetic and
self-interaction energy terms of a galactic halo described by a
Boehmer-Harko density profile. We discuss the validity of the
Thomas-Fermi approximation for the halo system, and show that the
kinetic energy contribution is indeed negligible.

\end{abstract}


\maketitle

\section{Introduction}

 In the search for a description of the dark matter that is
 responsible for most of the matter density in galaxies, many kinds of
 particles have been proposed. Among the most popular, we can cite
 WIMP's, {\it Weakly Interacting Massive Particles} \cite{pdg12},
 hypothetical particles with large masses that are being sought by
 many experiments. The mass proposed for this kind of particle lies in
 the range $10-100\; GeV$.

 The axion, a spin-0 particle with sub-eV mass proposed in the context
 of the Peccei-Quinn mechanism \cite{peccei}, is also considered a
 candidate for the galactic dark matter, in the form of a
 Bose-Einstein condensate (BEC) \cite{sik09} at low temperatures.

 Recently, it has been shown that a halo composed of an axionic
 Bose-Einstein condensate may present rotation curves that are a good
 fit for several galaxies \cite{boh07}.  This required the proposal of
 a specific density profile that we shall call Boehmer-Harko (BH)
 density profile.

 In \cite{pir12}, the radius of a halo composed of a condensate of
 spin-0 and spin-1 particles has been derived, and a statistical
 analysis has been performed to constrain the range of scattering
 lengths related to a particle in the axion mass range
 ($10^{-6}-10^{-4}\; eV$). This study has utilised the Boehmer-Harko
 density profile.

 One question arises in the application of such density profile to
 real galaxies, that of the halo stability. It has been claimed
 \cite{guz13} that the Boehmer-Harko density profile, obtained in the
 framework of the so-called Thomas-Fermi (TF) approximation, leads to the
 formation of a halo with positive total energy. Consequently, the
 halo is necessarily unstable. This conclusion takes into account that
 the particle is ultralight, with a mass $m=10^{-23}\; eV$ and the
 self-interaction is very weak, i.e, the scattering length is
 $a=10^{-80}\; m$.

 In the present work we argue that the Thomas-Fermi approximation is
 inescapable, due to the large number of particles in the halo. We
 also show, by estimating the kinetic energy contribution, that the
 total energy may be negative. Hence, the halo may be stable for the
 Boehmer-Harko density profile, given that the mass and scattering
 length of the axionic particle are appropriate.

 The plan of this paper is as follows. The section \ref{BEC}
 recapitulates the theoretical framework of Bose-Einstein condensation
 in the galactic case. The section \ref{num} presents a calculation of
 the total energy for a halo with a Boehmer-Harko density profile, and
 also gives a justification of the use of the Thomas-Fermi
 approximation. In section \ref{kin} the kinetic energy is calculated
 and compared to the interaction energy, along with a discussion on
 the halo stability. Finally, the section \ref{conc} presents our
 conclusions and final remarks.

\section{Bose-Einstein condensate halo}\label{BEC}

We recall here the theoretical description of the galactic
Bose-Einstein condensate composed of axionlike particles. 

We consider that each axionlike particle is represented by the field
destruction and creation operators, $\hat{\psi}({\bf r}, t)$ and
$\hat{\psi}^\dagger({\bf r},t)$. These field operators satisfy simple
commutation relation, $[\hat{\psi}({\bf r,t}),\hat{\psi}^\dagger({\bf
    r}',t)]=\delta({\bf r}-{\bf r}')$. The particles are
nonrelativistic, confined by the self-graviting potential, and only
two-body collisions with small momentum transfers play an important
role. Thus, the Hamiltonian operator is 
\begin{eqnarray}
\label{hamil}
\hat{H}=\int d^{3}{\bf
  r}\left[\hat{\psi}^\dagger({\bf r},t)\left(-\frac{\hslash^2}{2m}\nabla^2+V({\bf r})\right)\hat{\psi}({\bf r},t)+\frac{1}{2}\frac{4\pi\hslash^2a}{m}\hat{\psi}^\dagger({\bf r},t)\hat{\psi}^\dagger({\bf r},t)\hat{\psi}({\bf r},t)\hat{\psi}({\bf r},t)\right]\; ,
\end{eqnarray}
\noindent where $m$ is the mass of the particle, $a$ is the $s$-wave scattering
length which characterizes the collision and the trapping potential,
$V({\bf r})$, is determined by the Poisson's equation,
\begin{eqnarray}\label{poisson}
\nabla^2V=4\pi G m\rho_{DM}\;  ,
\end{eqnarray}
\noindent where $\rho_{DM}$ is the mass density of the dark matter halo.

At a sufficiently low temperature, there is a macroscopic occupation of $N_0$
particles in the lowest energy mode. Although the number operator
$\hat{N}=\int
\rd^{3}{{\bf r}}\hat{\psi}^\dagger({\bf r},t)\hat{\psi}({\bf r},t)$ commutes
with the hamiltonian (\ref{hamil}), the $U(1)$ symmetry of particle
number conservation is broken due to the ground state possessing a
coherent state $|\psi\rangle$ normalized to $N_0$ known as BEC
wavefunction.  We introduce this symmetry breaking through the Bogoliubov
replacement where the field operators are shifted by the BEC
wavefunction, $\psi({\bf r})$,
\begin{eqnarray}
\label{bog}
\hat{\psi}({\bf r},t)=e^{-i\mu
  t/\hslash}(\psi({\bf r})+\hat{\delta}({\bf r}))\; ,
\end{eqnarray}
\noindent where $\mu$ is the chemical potential.

The total number of particles $N$ is determined by the condition,
$N=N_0+\int \rd^{3}{{\bf r}}\langle\hat{\delta}^\dagger({\bf
  r})\hat{\delta}({\bf r})\rangle$, where the brackets denote the
ground state expectation values. For a dilute bosonic gas, it is
reasonable to consider $N-N_0\ll N$ and the Hamiltonian (\ref{hamil})
can be truncated to first order of $\hat{\delta}({\bf r})$ and
$\hat{\delta}^\dagger({\bf r})$.

At zero temperature, the dynamics of the field destruction operator
$\hat{\psi}({\bf r},t)$
in the Heisenberg picture,
$-i\hslash\partial_t\hat{\psi}({\bf r},t)=[\hat{H},\hat{\psi}({\bf r},t)]$, yields the
time-independent Gross-Pitaevskii equation (GPE) for the BEC wavefunction
$\psi(\vec{r})$
\be\label{GPEI}
\mu \psi({\bf r})=-\Frac{\hslash^2}{2m}\nabla^{2}\psi+V({\bf
  r})\psi({\bf r})+\Frac{4\pi \hslash^2 a}{m}|\psi({\bf
  r})|^{2}\psi({\bf r})\; .
\ee

It has been demonstrated \cite{boh07, pir12} that (\ref{GPEI}) has
the solution 
\be\label{psibh}
\psi_{BH}(r)=\begin{cases} \sqrt{\rho_0\Frac{\sin
    kr}{kr}}\quad\mbox{for}\quad r\le R \\  0\quad \mbox{for} \quad
r>R \end{cases}\; ,
\ee
\noindent with $k=\sqrt{Gm^3/\hslash^2a}$ and $R=\pi/k$. $\rho_{0}$ is
the central particle number density of the condensate. This is the
Boehmer-Harko solution in the Thomas-Fermi approximation. It results
in a halo radius given by 
\be\label{radius}
R=\pi \sqrt{\Frac{\hslash^{2} a}{G m^{3}}}\;.
\ee
\noindent Using this relation and considering the dark matter particle mass
range $10^{-6}-10^{-4}\; eV$, the lower bound of the scattering length has been
constrained to $10^{-29}\ m$ \cite{pir12}.

The central mass density $\rho_{DM}=m\rho_{0}$ will be assumed throughout
this paper to be of the same order of magnitude of the local dark
matter density in the Milky Way, $\rho=0.47\; GeV/cm^{3}$
\cite{nesti13}.

\section{The total energy of a halo with Boehmer-Harko density profile}\label{num}

The zero-temperature mean field energy of a weakly interacting BEC
confined in a self-graviting potential, $V$, is given by \cite{pir12}
\begin{eqnarray}\label{energy}
E=\langle\hat{H}\rangle=K+W+I\; ,
\end{eqnarray}
\noindent where  the kinetic,
potential and self-interaction energies, respectively, are
\begin{eqnarray}
K &=& \int \rd^{3}{\bf
  r}\; \psi^*\left(-\frac{\hslash^2}{2m}\nabla^2\right)\psi\; ,\\
W &=& \int \rd^{3}{\bf r}\; \psi^*V\psi\; ,\\
I &=& \frac{2\pi\hslash^2a}{m}\int \rd^{3}{\bf
  r}\;|\psi|^4\;.
\end{eqnarray}

 The expression for the kinetic energy above is also called quantum
 pressure, and it can be shown to be negligible for a large total
 number of particles. In fact, this is essential for the so-called
 Thomas-Fermi approximation, which is the large $N$ limit of the
 solution for the density of the condensate. It is useful to compare
 the gravitationally bounded system we are investigating with a
 trapped condensate that can be obtained in laboratory.

 For a Bose-Einstein condensate confined by a harmonic oscillator
 trap with a potential $V(r)=\frac{m\omega_{ho}^{2}}{2}r^{2}$, where
 $\omega_{ho}$ is the trap frequency, it is possible to show that the
 TF approximation is valid for $\Frac{Na}{a_{ho}}\gg 1$ \cite{str99},
 where 
\be
a_{ho}=\left(\Frac{\hslash}{m\omega_{ho}}\right)^{1/2}
\ee

 \noindent is the harmonic oscillator length, $N$ is the total particle number
 and $a$ is the scattering length implementing the interaction between
 the atoms in the condensate. The radius of the condensate in the
 limit in which TF approximation is valid is 
\be
R=a_{ho}^{4/5}(15 N a )^{1/5}\;.  
\ee

Hence, in terms of the TF radius, we have
\be 
\Frac{15(Na)^5}{R^5}\gg 1\; .
\ee

This limit is easily achievable in laboratory since the condensate is
composed of $N\sim 10^{6}$ atoms.

Now, making an analogy with the laboratory case, the same condition
for the TF approximation must hold, i. e., $\Frac{Na}{R}\gg 1$. Of
course, the parameters refer now to the galactic
features. Considering the typical values $N=10^{82}$,
$a=10^{-29}\; m$ and $R=10^{20}\; m$ \cite{pir12}, we have
$\Frac{Na}{R}=10^{33}\gg 1$. Therefore, the TF approximation
is always valid in the galactic condensate case.

An important consequence of this approximation is that we can neglect
the quantum pressure term in the energy expression, because the
relative contribution of the kinetic energy of the particles becomes
smaller in comparison to the interaction energy when the
number of particles increases. The total energy of the condensate
is represented now by $E=W+I$.

\begin{figure}[!htp]
\begin{center}

\includegraphics[scale=0.7]{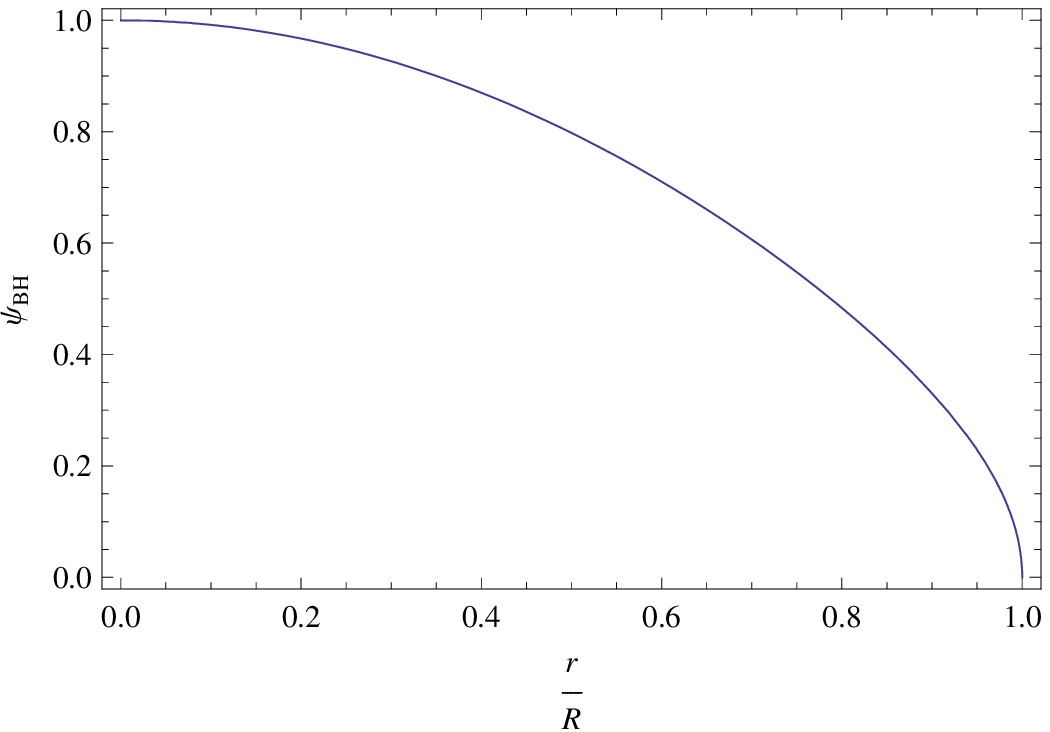}
\hspace{0.3cm}
\includegraphics[scale=0.7]{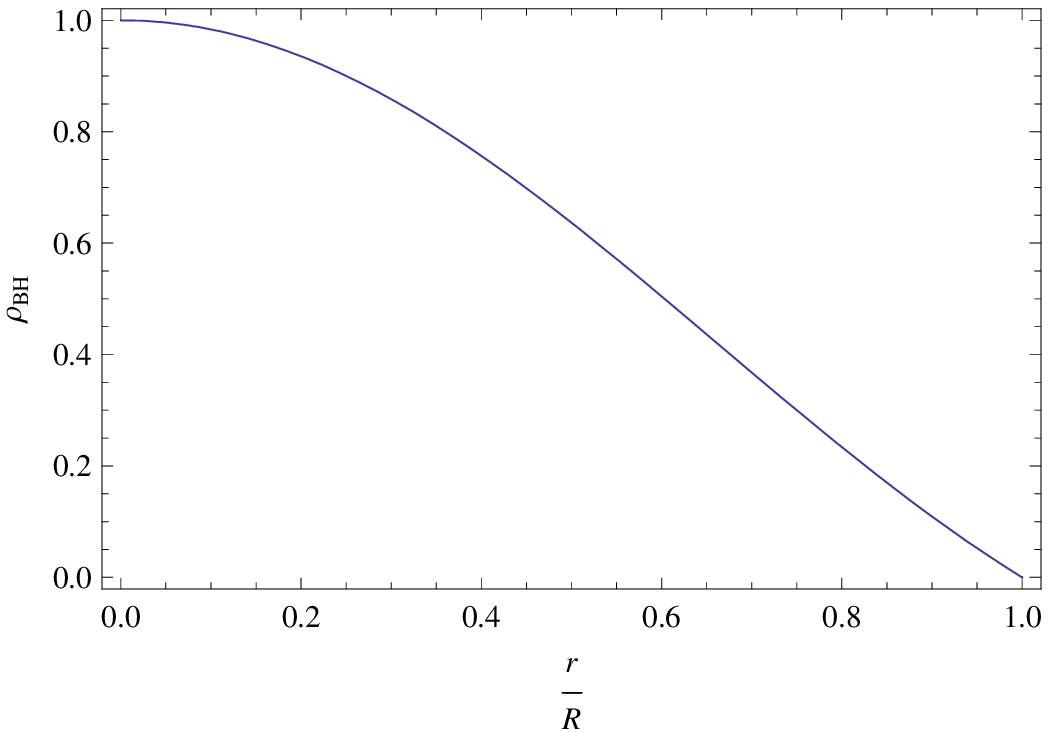}

\caption{Wavefunction $\psi_{BH}$ (in units of
  $\sqrt{(\rho_{0}/\pi)}$) and density profile $\rho_{BH}$ (in units of
  $(\rho_{0}/\pi)$) corresponding to the Boehmer-Harko solution in the
  Thomas-Fermi approximation.}
\label{bh_solution}

\end{center}
\end{figure}

Using the solution given by \cite{boh07} for the density profile
in terms of $|\psi_{BH}(r)|^{2}$ in the TF approximation, we have
\begin{eqnarray}
\rho_{BH}=|\psi_{BH}(r)|^{2}=\begin{cases} \rho_0\Frac{\sin
    kr}{kr}\quad\mbox{for}\quad r\le R \\  0\quad \mbox{for} \quad
r>R \end{cases}\; .
\end{eqnarray}
\noindent Both the wavefunction and
the density profile for the condensate are depicted in figure
\ref{bh_solution}.

With this solution, we can calculate the interaction energy as
\be\label{inter} 
I=\frac{2\pi\hslash^2a}{m}\int \rd^{3}{\bf
  r}\;|\psi|^4=4\Frac{\hslash^{2}a}{m}\rho_{0}^{2}R^{3} \; .
\ee

We can see that this quantity is always positive, as long as the
interaction is repulsive ($a>0$).

For a radial homogeneous distribution the mass inside a radius $r$ can
be computed as 
\be 
M(r)=4\pi \int_{0}^{r}m|\psi|^{2}r'^{2}\rd r' \; .
\ee

Considering the BH density profile, we can write
\begin{eqnarray}\label{mass}
M(r)=\frac{4\pi Gm\rho_0 r}{k^2}\left(\frac{\sin (kr)}{kr}-\cos
(kr)\right)\; .
\end{eqnarray}

The total mass of the halo is $M_{T}=\Frac{4}{\pi}m\rho_{0}R^{3}$.

The gravitational field for such radial mass distribution is
given by
\begin{eqnarray}
g(r)=\begin{cases} -\Frac{GM(r)}{r^{2}}=-\Frac{4\pi Gm\rho_0 }{k^2 r}\left(\Frac{\sin (kr)}{kr}-\cos
(kr)\right)\quad\mbox{for}\quad r < R \\ 
-\Frac{GM_{T}}{r^2}=-\Frac{4}{\pi}Gm\rho_{0}\Frac{R^{3}}{r^{2}}\quad \mbox{for} \quad
r\geqslant R \end{cases}\; .
\end{eqnarray}

Hence, we find that the gravitational potential $V(r)$ is
\be\label{gravpot}
V(r)=m \int^{\infty}_{r}g(r')\rd r'=-\Frac{4\pi m^{2}G}{k^{2}}|\psi|^{2}-\Frac{4}{\pi}m^{2}
G\rho_{0}R^{2}\; .
\ee

Now we can proceed to calculate the gravitational potential
energy of the condensate
\be\label{gravenerg}
W=\int_{0}^{R} V(r)|\psi(r)|^{2}\rd^{3}r=-4\left(4 \Frac{\hslash
  a}{m}\rho_{0}^{2}R^{3}\right)=-4I \;.
\ee

The total energy becomes $E=W+I=-3I$. 

The validity of this approach is shown in the next section.

\section{Kinetic energy and halo stability}\label{kin}

Computing the kinetic energy in the TF approximation, inside
a halo with radius $R$, we find the expression
\be\label{kinetic}
K=\Frac{\pi\hslash^{2}\rho_0}{4
  m}\left[\int_{0}^{R}|\Psi|^{2}\left(2k^{2}+k^{2}\cot^{2}(k
  r)-\Frac{2k\cot(kr)}{r}+\Frac{1}{r^2}\right)r^2 \rd r\right]\; ,
\ee
\noindent which is non-convergent in the interval $[0, R]$. We will
refer to the integral inside the brackets in (\ref{kinetic}) as $\mathcal{I}$.
\begin{figure}[!htp]
\begin{center}
\includegraphics[scale=1.0]{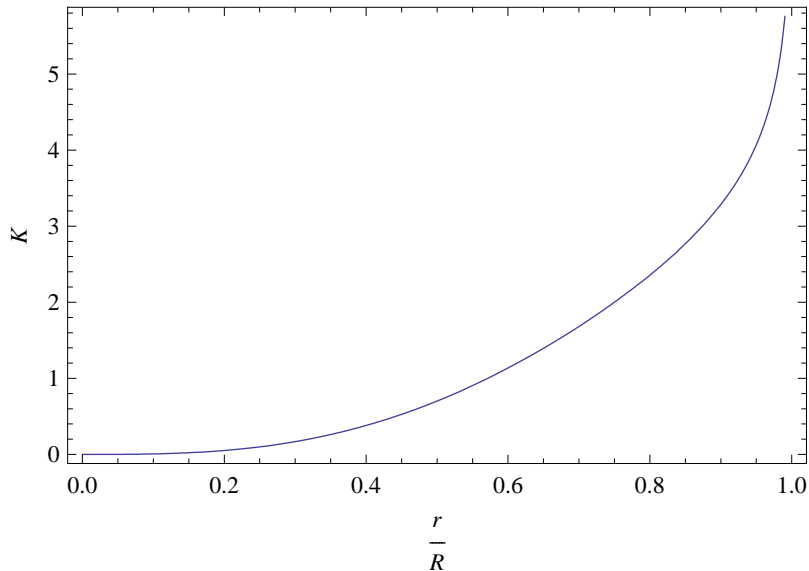}
\caption{The kinetic energy (in units of $(\pi\hslash^{2}\rho_0 R)/(4
  m)$) as a function of the dimensionless radius $r/R$. There
  is a logarithmic divergence as $r/R \rightarrow 1$. The plot is made
  for the interval $[0,(1-10^{-7}) R]$, avoiding this divergence.}
\label{kineticfig}
\end{center}
\end{figure}
  
 It can be shown that this quantity can be estimated as
 \be\label{int_kin} 
K=\Frac{\pi\hslash^{2}\rho_0R}{4
   m}\left[Si(\pi)-\pi+\lim_{x \to \pi}\left(x \ln
   \left(\tan\left(\frac{x}{2}\right)\right)\right)\right]\; , 
\ee
\noindent where $x=\Frac{\pi r }{R}$.

 The last term inside the brackets in (\ref{int_kin}) shows a
 logarithmic divergence (see figure \ref{kineticfig}). This term
 cannot be exactly calculated up to $R$, but up to a value that is
 close enough to represent the halo interior it results in a small
 numerical value. The issue of the calculation of the kinetic energy
 as a border effect beyond the TF approximation has been treated in
 the context of atomic condensates \cite{lun97, dal96, fet98}, and it
 is beyond the scope of the present work. Nevertheless, it is worthy
 to mention that these investigations have found that the kinect
 energy contribution in a region very close to the border of the
 condensate is negligible, even for systems with as few as $10^{5}$
 particles.

 In fact, numerical integration of (\ref{kinetic}) (up to
 $r=(1-10^{-16})R$) results in $\mathcal{I} \approx 38$. Hence, the
 order of magnitude of the kinetic energy is mainly given by the
 multiplying factor $\frac{\pi\hslash^{2}\rho_0 R}{4 m}$.

 In order to establish the relative importance of the kinetic energy
 term in relation to the total condensate energy, we can calculate the
 ratio $\eta$ of the kinetic energy to the interaction energy 
\be\label{ratio} 
\eta=\Frac{K}{3I}\sim \Frac{1}{a \rho_0 R^{2}}\; .
\ee
 If $\eta \leqslant 1$, the total energy is negative, and if $\eta>1$ the
 kinetic energy should be large enough to make the total energy
 positive. 

 For the cases shown in \cite{guz13}, with the values
 $m=10^{-22}\, eV$, $a\approx 10^{-80}\, m$ and $R\approx 10^{20}\,
 m$, this ratio results in 
\be
\eta\approx 10^{4}.
\ee
 \noindent This seems to be the reason why the authors in \cite{guz13}
 obtained a positive total energy for the condensate in the TF
 approximation, and concluded that the halo described by the BF
 density profile is necessarily unstable. Also, such a small
 scattering length implies an almost null interaction parameter,
 justifying the use by the authors of the Gaussian approximation
 for the density profile instead of the TF approximation.

 We want to emphasize that this result is strongly dependent on the
 values chosen for the quantities $m$ and $a$, and different masses
 and scattering lengths of the dark matter particle can lead to
 different conclusions about the halo stability. For
 instance, choosing $m=10^{-6}\, eV$ and $a=10^{-29}\, m$ \cite{pir12},
 we obtain
\be
\eta \approx 10^{-31}\; ,
\ee
\noindent i.e., the kinetic energy is really negligible, and the total
energy is negative, showing that the halo endowed with a Boehmer-Harko
density profile can be stable in this specific case.

\begin{figure}[!htp]
\begin{center}
\includegraphics[scale=0.8]{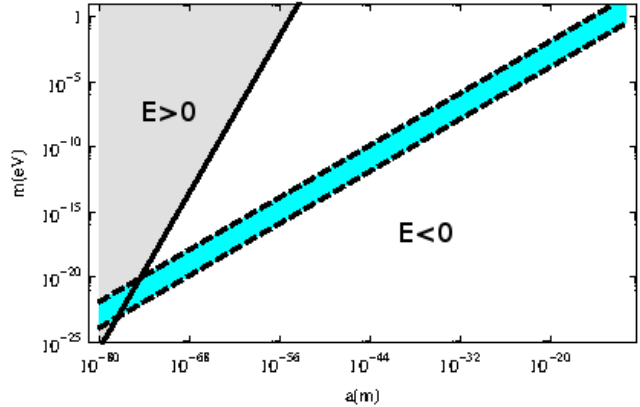}
\caption{Parameter space ($m$,$a$) for the condensate model. The
  shaded area is the region in which the total energy is positive. In
  the white area the energy is negative. The straight line is $\eta=1$
  for $R=10^{20}\; m$. The blue area between the dashed lines is the
  region where the values of the parameters $m$ and $a$ result in the
  galactic radii range $0.1-10\; kpc$ \cite{pir12} (color online).}
\label{paramspace}
\end{center}
\end{figure}

 The figure \ref{paramspace} shows the parameter space for the
 condensate model with a BH profile. We can see that most of it allows for negative
 energy. The choice of mass range $10^{-6}\ eV<m<10^{-4}\ eV$, which
 has been determined in \cite{pir12} for axionlike particles, and
 resulting in galaxies' radii ranging from $\sim 0.1\; kpc$ to $\sim 10\;
 kpc$, is shown as the blue area in this plot.

 A few words must be said about the criticism on the BH solution
 inability to yield different halo radii. The halo radius derived from
 this solution is a function of the fundamental dark matter quantities
 $m$ and $a$, and, as a consequence, it represents a prototypical
 fundamental dark matter halo. Once a particle mass has been chosen,
 the only free parameter allowing to obtain different radii is the
 scattering length. As we can see from the plot \ref{paramspace}, in
 order to reflect observed galactic radii this choice is constrained
 in the parameter space.

 The main difference between the BH profile and other profiles used to
 study galactic dynamics is that BH is obtained from first principles
 of Quantum Physics, while the other ones are usually
 phenomenologically fit functions. This fact may be important in the
 investigation of microscopic properties of the dark matter particles.


 In principle, the results presented here are applicable to the galaxy
 cluster range, at least as a phenomenological
 approximation. However, we believe that the microscopic description,
 involving Gross-Pitaevskii equation, for instance, should require a
 more precise relativistic treatment for the cluster case.


 We remark that, in the cluster scale, cosmological parameters
 regarding the Universe's expansion are relevant for the study of the
 system's dynamics and the determination of its energy content. This
 can be implemented, for example, by the use of the Layzer-Irvine
 equation for the evolution of the energy of cold dark matter in an
 expanding environment \cite{peebles}. The relativistic approach for
 the problem is not the aim of the present work.

\section{Conclusions}\label{conc}

 We have given expressions for the kinetic ($K$), potential ($W$) and
 self-interaction ($I$) energy components of the Bose-Einstein
 condensate dark matter halo composed of axionlike particles and
 described by the Boehmer-Harko density profile. These quantities have
 been defined by the fundamental parameters of the condensate (the
 particle mass $m$ and the scattering length $a$ of the interaction)
 and the halo parameters (central density $\rho_{0}$ and radius
 $R$). We have found that the total energy $E=W+K+I$ may be written
 $E=W+I$ in the Thomas-Fermi approximation. Moreover, we have found
 that $W=-4I$, rendering the total energy negative. By comparing the
 conditions for validity of this approximation in atomic condensates
 created in laboratory with the axionlike halo system, we have found
 that $\Frac{Na}{R}\gg 1$, where $N$ is the total particle
 number. This condition is always fulfilled in the galactic case since
 $N\sim 10^{82}$ and $10^{-6}\; eV< m < 10^{-4}\; eV$. Hence, the TF
 approximation should be valid for the BH density profile.

 In order to stress this fact and to show the strong dependence of the
 energy terms expressions on the mass and scattering length of the
 dark matter particle, we have performed a semi-analytical calculation
 of the kinetic energy term, and showed that the ratio $\eta=K/3I$
 indicates the sign of the total energy. For the case of the axionlike
 particle with mass $m=10^{-6}\, eV$ \cite{pir12}, we have $\eta
 \approx 10^{-31}$. Therefore, the total energy in this case is
 negative and the system should be stable.

 On the other hand, for $m=10^{-23}\, eV$ and $a= 10^{-80}\; m$
 \cite{guz13}, $\eta\approx 10^{4}$ and the system is unstable. We
 point out that the choice of such a small scattering length makes the
 particle interaction and the potential energy negligible. 

 Since the total energy is so sensitive to the values of mass and
 scattering length we claim that its magnitude alone should not be
 used to ascertain the stability of the system described by the BH
 density profile nor to rule out the validity of the TF
 approximation. Rather, we believe that it is necessary to find
 methods to obtain the values of $m$ and $a$ independently and then
 resort to procedures as phase space analysis and numerical solutions
 of the differential equations involved. This search will be the
 subject of future work.

 We want to emphasize that we use Bose-Einstein condensation as an
 analog for a dark matter halo, taking advantage of the possibility of
 obtaining experimental information in a controled
 manner. Nevertheless, the fact that the BH profile is capable of
 satisfactorily fitting rotation curves for a number of galaxies may
 be an indication of its adequacy in describing the underlying
 galactic dynamics.

\begin{acknowledgments}
 J. C. C. S. thanks CAPES (Coordena\c c\~ao de Aperfei\c coamento de
 Pessoal de N\'\i vel Superior) for financial support. The authors
 wish to thank an anonymous referee for useful suggestions.
\end{acknowledgments}

\end{document}